\documentclass[prb,twocolumn,superscriptaddress,preprintnumbers]{revtex4}

\usepackage{graphicx}
\usepackage{dcolumn}
\usepackage{amsmath}
\usepackage{color}
\usepackage[utf8]{inputenc}

\newcommand{\beginsupplement}{%
        \setcounter{table}{0}
        \renewcommand{\thetable}{S\arabic{table}}%
        \setcounter{figure}{0}
        \renewcommand{\thefigure}{S\arabic{figure}}%
}

\begin{document}

\title{Electronic correlations at paramagnetic $(001)$ and $(110)$ NiO surfaces: Charge-transfer and Mott-Hubbard-type gaps at the surface and subsurface of $(110)$ NiO}

\author{I. Leonov}
\affiliation{M. N. Miheev Institute of Metal Physics, Russian Academy of Sciences, 620108 Yekaterinburg, Russia}
\affiliation{Ural Federal University, 620002 Yekaterinburg, Russia}
\affiliation{Skolkovo Institute of Science and Technology, 143026 Moscow, Russia}

\author{S. Biermann}

\affiliation{CPHT, CNRS, Ecole Polytechnique, IP Paris, F-91128 Palaiseau, France}
\affiliation{Coll\' ege de France, 11 place Marcelin Berthelot, 75005 Paris, France}
\affiliation{Department of Physics, Division of Mathematical Physics, Lund University, Professorsgatan 1, 22363 Lund, Sweden}

\begin{abstract}

We explore the interplay of electron-electron correlations and surface effects in the prototypical correlated insulating material, NiO. In particular, we compute the electronic structure, magnetic properties, and surface energies of the $(001)$ and $(110)$ surfaces of paramagnetic NiO using a fully charge self-consistent DFT+dynamical mean-field theory method. Our results reveal a complex interplay between electronic correlations and surface effects in NiO, with the electronic structure of the $(001)$ and $(110)$ NiO surfaces being significantly different from that in bulk NiO. We obtain a sizeable reduction of the band gap at the surface of NiO, which is most significant for the $(110)$ NiO surface. This suggests a higher catalytic activity of the $(110)$ NiO surface than that of the $(001)$ NiO one. Our results reveal a charge-transfer character of the $(001)$ and $(110)$ surfaces of NiO. Most notably, for the $(110)$ NiO surface we observe a remarkable electronic state characterized by an alternating charge-transfer and Mott-Hubbard character of the band gap in the surface and subsurface NiO layers, respectively. This novel form of electronic order stabilized by strong correlations is not driven by lattice reconstructions but of purely electronic origin. We notice the importance of orbital-differentiation of the Ni $e_g$ states to characterize the Mott-Hubbard insulating state of the $(001)$ and $(110)$ NiO surfaces. The unoccupied Ni $e_g$ surface states are seen to split from the lower edge of the conduction band to form strongly localized states in the fundamental gap of bulk NiO. Our results for the surface energies of the $(001)$ and $(110)$ NiO surfaces show that the $(001)$ facet of NiO has significantly lower energy. This implies that the relative stability of different surfaces, at least from a purely energetic point of view, does not depend on the presence or absence of magnetic order in NiO.
  
\end{abstract}
\maketitle

\section*{I. INTRODUCTION}

The series of transition metal monoxides MnO, FeO, CoO, and NiO with an electronic configuration ranging from $3d^5$ to $3d^8$, respectively, has attracted much attention due to their diverse electronic and magnetic properties \cite{Noguera_1996,Henrich_1994,Kung_1989,Freund_1996,Kaiser_2007}, allowing for a broad range of applications, e.g., in electronics and spintronics \cite{Liu_2018,Lin_2016,Aytan_2017,Dabrowski_2020}, energy storage \cite{Poizot_2000,Reddy_2013,Su_2012,Lee_2016}, and heterogeneous catalysis \cite{Zhao_2016,Nelson_2016, Poulain_2018,Gong_2020}. At low temperature, these compounds exhibit a correlated Mott-Hubbard or charge-transfer insulating behavior with a large band gap of $\sim$2-4 eV, associated with a strong localization of the $3d$ electrons \cite{Mott_1990,Imada_1998,Sawatzky_1984,Zaanen_1985,Shen_1991}. Below the N\'eel temperature, ranging from $T_N \sim 116$ to 523 K for MnO to NiO, respectively, these materials display an antiferromagnetic type-II long-range magnetic ordering and undergo a structural phase transition from a rocksalt cubic to a distorted rhombohedral (MnO and NiO) or monoclinic (FeO and CoO) phase \cite{Roth_1958}. 

Over the past decades particular attention has been devoted to understanding the nature of the band gap and excitation spectrum of MnO-NiO \cite{Mott_1990,Imada_1998,Sawatzky_1984,Zaanen_1985,Shen_1991,Schuler_2005,Anisimov_1991,Dudarev_1998,Cococcioni_2005,Zunger-review,Wang_2020,Potapkin_2016}.
In fact, due to the strongly correlated nature of electron interactions between the $3d$ electrons, theoretical computations of the electronic structure of these materials using band-structure methods are particularly difficult. The Coulomb interactions may be modelled by including an on-site Hubbard parameter $U$ to treat the effect of correlations in the partially filled $3d$ shell, e.g., within the so-called DFT+$U$ method \cite{Anisimov_1991,Dudarev_1998,Cococcioni_2005}. While these ``beyond standard density functional theory (DFT)'' methods often give a reliable description of the electronic properties such as band gaps, magnetic moments, and lattice displacements \cite{Potapkin_2016,Schron_2012,Rodl_2012,Jiang_2010,Karlsson_2010,Liu_2020},
these methods neglect electron dynamics and hence cannot capture correlated electron phenomena related to a Mott transition such as a coherence-incoherence crossover, quasiparticle behavior, and strong quasiparticle mass renormalization \cite{Mott_1990,Imada_1998,DMFT-review,Kotliar_review,Springer-Review}. In fact, to determine the electronic properties of these materials one needs to go beyond conventional band-structure methods to include dynamical correlation effects of the $3d$ electron (e.g., using DFT+$U$ subject to dynamical symmetry-breaking spin and structural effects \cite{Zunger-review,Wang_2020}).  

It has been shown that the dynamical correlations and strong localization of the $d$ (and $f$) electrons can be described by employing a DFT+dynamical mean-field theory (DFT+DMFT) approach \cite{DMFT-review,Kotliar_review,Springer-Review}. DFT+DMFT makes it possible to treat local correlation effects in a self-consistent, numerically exact way, providing a good description of the electronic and magnetic properties of correlated models and materials \cite{DMFT-review,Kotliar_review,Springer-Review,Mn,Pavarini_2004,Ti2O3,VO2,Kunes_2007,Kunes_2008,Yin_2008,Shorikov_2010,Ohta_2012,Thunstrom_2012,Byczuk_2012,Nekrasov_2012,Leonov_2015,Panda_2016,Leonov_2016,Greenberg_2020,Panda_2017,Lechermann_2019,
Mandal_2019,Lanata_2019}. By using DFT+DMFT it becomes possible to compute the material-specific properties of complex correlated materials, e.g., to determine the electronic structure, magnetic state, and crystal structure of (para-) magnetic materials at finite temperatures, e.g., near the Mott transition \cite{Kunes_2008,Shorikov_2010,Ohta_2012,Leonov_2015,Leonov_2016,Greenberg_2020,Lanata_2019,Leonov_2018}.

The DFT+DMFT approach has been widely used to study the physical properties of the bulk structure of transition metal monoxides (TMOs), providing a quantitative description of the electronic, magnetic, and structural properties of these materials \cite{Kunes_2007,Kunes_2008,Yin_2008,Shorikov_2010,Ohta_2012,
Thunstrom_2012,Byczuk_2012,Nekrasov_2012,Leonov_2015,Panda_2016,
Leonov_2016,Greenberg_2020,Panda_2017,Lechermann_2019,Mandal_2019,Lanata_2019}. While the electronic structure and magnetic state of bulk TMOs are relatively well established nowadays (see however \cite{Karolak,Tjeng-DFG,Tjeng_2017}),
this is not the case for the different surface structures of metal oxides and their interfaces \cite{Noguera_1996,Henrich_1994,Kung_1989,Freund_1996,Kaiser_2007}. In fact, the electronic structure of a surface or interface can differ significantly from that of bulk, since their electronic states are responsible for a variety of novel (emerging) physical properties. In the case of surfaces, this affects, e.g., molecular adsorption and reactions that determine the catalytic processes at the surface \cite{Noguera_1996,Henrich_1994,Kung_1989}. In this respect, NiO is of particular interest among other TMOs due to its importance as catalysts for electrochemical applications, fuel cells, and batteries \cite{Poizot_2000,Reddy_2013,Su_2012,Lee_2016,Zhao_2016,Nelson_2016, Poulain_2018,Gong_2020} (in addition to being a model system for understanding the Mott-Hubbard metal-insulator transition \cite{Mott_1990,Imada_1998}). 

The electronic properties of the nonpolar $(001)$ and $(110)$, and polar $(111)$ NiO surfaces have been actively studied using different experimental techniques, such as x-ray photoemission spectroscopy (XPS), x-ray linear magnetic dichroism, electron-energy-loss spectroscopy (EELS), and scanning tunnelling microscopy (STM), that give a detailed information about the surface $d$-shell excitations and atomically resolved images of the surface states and reconstructions \cite{Noguera_1996,Henrich_1994,Kung_1989,Freund_1996,Kaiser_2007,Fromme_1996,Alders_1998,Stohr_1999,Barbier_2000,Ohldag_2001,Hillebrecht_2001,Soriano_2007,Pielmeier_2013}. On the other hand, the effect of electron correlations and the electronic structure and magnetic properties of the NiO surfaces are still poorly understood. In practice, the electronic and magnetic properties of NiO surfaces have been studied using band-structure methods with the static mean-field Hubbard $U$ treatment of correlation effects in the Ni $3d$ shell (e.g., within DFT+$U$) \cite{Castell_1999,Hoeft_2001,Kodderitzsch_2002,Bengone_2002,Wander_2003,Rohrbach_2004,Ferrari_2007,Yu_2008,Schron_2013,Chaitanya_2020}. The strong localization of the $3d$ electrons and the finite temperature (para-) magnetic behavior of NiO still pose a challenge for an accurate description of the electronic properties of its surfaces using, e.g, DFT+DMFT \cite{Potthoff_1999,Potthoff_2002}, which is crucial for potential technical applications \cite{Poizot_2000,Reddy_2013,Su_2012,Lee_2016,Zhao_2016,Nelson_2016, Poulain_2018,Gong_2020}.  

In this paper, we employ a multi-site extension of the DFT+DMFT method \cite{vanRoekeghem, Delange,Ti2O3,VO2} implemented within a plane-wave pseudopotential formalism \cite{Leonov_2015,Leonov_2016,Greenberg_2020,Leonov_2018} to explore the effects of electronic correlations on the electronic properties of the nonpolar paramagnetic (PM) $(001)$ and $(110)$ surfaces of NiO. We study the electronic structure, magnetic properties, and surface energies of the $(001)$ and $(110)$ NiO with a particular attention given to the effect of structural confinement and its influence on the strength of electronic correlations in the PM $(001)$ and $(110)$ NiO. 

\section*{II. COMPUTATIONAL DETAILS}

We perform a DFT+DMFT study of the electronic structure, magnetic properties, and surface energies of the clean $(001)$ and $(110)$ surfaces of the prototypical correlated insulator NiO. In DFT we employ the generalized gradient Perdew-Burke-Ernzerhof approximation (GGA) \cite{Perdew_1996} within plane-wave  pseudopotential formalism \cite{pseudo}. Our DFT+DMFT calculations explicitly include the Ni $3d$ and O $2p$ valence states, by constructing a basis set of atomic-centered Wannier functions within the energy window spanned by the $p$-$d$ band complex \cite{Wannier}. This makes it possible to take into account a charge transfer between the Ni $3d$ and O $2p$ states, accompanied by the strong on-site Coulomb correlations in the Ni $3d$ shell. In order to solve the realistic many-body problem within DMFT we use the continuous-time hybridization-expansion quantum Monte-Carlo algorithm \cite{ctqmc}. 

The DFT+DMFT calculations are performed in the \emph{paramagnetic} state at an electronic temperature $T = 390$ K. We use the average Hubbard $U = 10$ eV and Hund's exchange $J = 1$ eV, in accordance with previous work \cite{Anisimov_1991,Leonov_2016,Panda_2017,Sakuma_2013}. The Coulomb interaction $U$ and Hund's $J$ have been treated in the density-density approximation, with no spin-orbit coupling contribution. We employ the fully localized double-counting correction evaluated from the self-consistently determined local occupations, to account for the electronic interactions already described by DFT. We use a fully self-consistent in charge density implementation of DFT+DMFT in order to take into account the effects of charge redistribution caused by electronic correlations and electron-lattice coupling \cite{charge-sc-LDA+DMFT,Leonov_2016}.  In order to study the effects of structural confinement and relaxations we employ a multi-impurity-site extension of the DFT+DMFT method in order to treat correlations in the $3d$ bands of the structurally distinct Ni sites \cite{vanRoekeghem, Delange,VO2,Ti2O3}. In addition, we perform structural relaxations of the surface and subsurface states of NiO within DFT. The DFT structural relaxation calculations are performed for the (fictitious) ferromagnetic (FM) state of NiO. For the relaxed structures the electronic properties of $(001)$ and $(110)$ PM NiO were evaluated within DFT+DMFT. The spectral functions were computed using the maximum entropy method and the Pad\'e analytical continuation procedure.

\section*{III. RESULTS AND DISCUSSION}

\subsection*{a. Electronic structure of $(001)$ NiO}

We begin with the electronic structure and equilibrium lattice volume calculations of bulk NiO in the paramagnetic state using DFT+DMFT \cite{Leonov_2015,Leonov_2016}. In agreement with experiment, our calculations yield a correlated insulator with a large Ni $d$-$d$ energy gap of about 2.9 eV and equilibrium lattice constant $a=4.233$ \AA\ (lattice volume of 128 a.u.$^3$ and bulk modulus 187 GPa). The Ni$^{2+}$ $d^8$ ions are in a high-spin $(S=1)$ state with an instantaneous local magnetic moment of 1.83 $\mu_\mathrm{B}$, corresponding to the fluctuating moment of 1.7 $\mu_\mathrm{B}$. This result is in good agreement with the experimental estimate of 1.7-1.9 $\mu_\mathrm{B}$ \cite{NiO_exp_moments}. 
We note that the top of the valence band of NiO has a mixed Ni $e_g$ and O $2p$ character, with a resonant peak in the filled Ni $e_{g}$ band located at about 1 eV below the Fermi level. The latter can be ascribed to the formation of a Zhang-Rice bound state \cite{Zhang_1988}. Our results suggest a mixture of a Mott-Hubbard Ni $d$-$d$ and charge transfer type of the band gap which is caused by the effect of Coulomb correlations in the Ni $3d$ shell \cite{Sawatzky_1984,Zaanen_1985,Schuler_2005}. The bottom of the conduction band is mainly of the Ni $3d$ character with a broad paraboliclike Ni $4s$ band at the Brillouin zone $\Gamma$ point.

Next, we use the equilibrium lattice constant obtained by DFT+DMFT for bulk NiO to construct the $(001)$ symmetric slab consisting of 7-NiO-monolayers (ML), with a thickness of about 12.59 \AA\ and vacuum of $\sim$21 \AA. Each ML in the slab contains two Ni sites. The top two NiO MLs are considered as a surface and subsurface, respectively, with 3-ML-thick NiO as a quasi-bulk. In our calculations we set the kinetic energy cutoff for the plane wave basis to 65 Ryd for the wave function and 650 Ryd for the charge density. We employ the three-impurity-site DFT+DMFT to treat structurally distinct the surface, subsurface, and quasi-bulk Ni $3d$ ions \cite{Keshavarz_2015}. We first perform structural optimization of the surface and subsurface $(001)$ NiO layers within FM GGA. It gives a small inward relaxation of the top ML of $\delta_{12} \simeq -2.5$\% (0.054 \AA) and a weak outward relaxation of the subsurface ML $\delta_{23} \sim 0.06$\%, with nearly absent surface buckling. Here, surface relaxations are characterized as a percent change of the spacing between layers $i$ and $j$ versus the equilibrium inter-layer spacing $d_0$,  $\delta_{ij}=(z^\mathrm{Ni}_i-z^\mathrm{Ni}_j)/d_0$. We also find a small inward inter-plane relaxation between the first and third MLs, $\delta_{13}\sim -1.0$\%. Interestingly, the nomagnetic GGA yields -2.7\% and -2.4\% relaxation for the top and subsurface NiO-MLs, respectively, with a small surface buckling of $\sim$0.11 \AA.

\begin{figure}[t]
\includegraphics[width=0.45\textwidth]{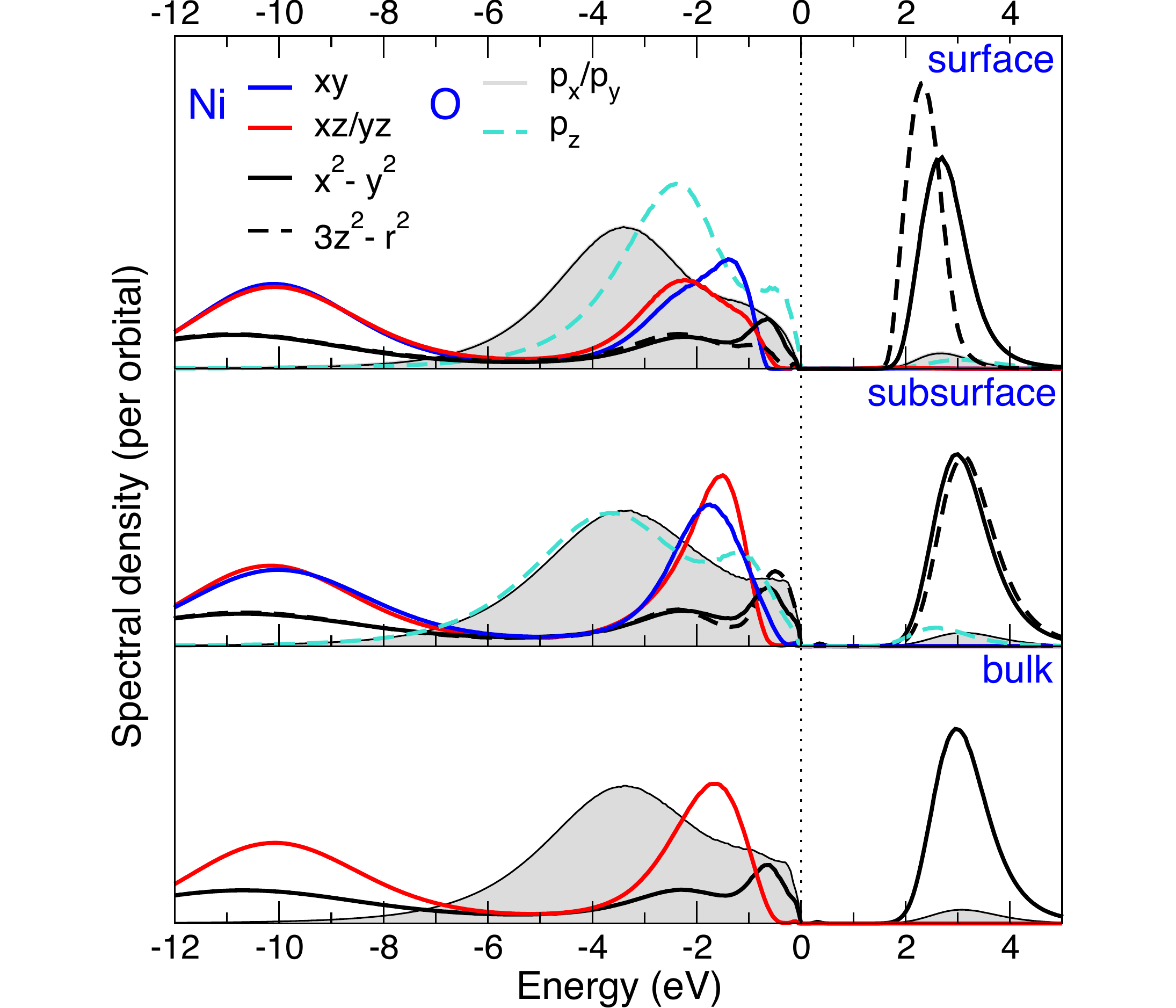}
\caption{(Color online) Ni $3d$ and O $2p$ spectral functions calculated by DFT+DMFT for the relaxed $(001)$ NiO at a temperature $T=390$~K.}
\label{Fig_1}
\end{figure}

In Fig.~\ref{Fig_1} we show the orbitally-resolved spectral functions of $(001)$ NiO obtained by DFT+DMFT for the relaxed within FM GGA surface of NiO. Our results for the unrelaxed (bulk-terminated) $(001)$ NiO are shown in Supplementary Fig. S1. In accordance with experiment, we find a Mott-Hubbard insulator with a large (fundamental) energy gap of about 2.1 eV. The calculated local magnetic moments are 1.84, 1.82, and 1.82 $\mu_B$ for the surface, subsurface, and quasi-bulk Ni ions, respectively. The corresponding fluctuating moments are 1.78, 1.74, and 1.75~$\mu_\mathrm{B}$. Our results reveal a weak charge redistribution between the surface and bulk Ni sites. The total Wannier Ni $3d$ occupancies are nearly the same for all the Ni sites and differ by less than 0.03. In addition, we find no sizeable distinction between the Ni $3z^2-r^2$ and $x^2-y^2$ orbital occupancies, which differ by less than 0.03 for all the Ni sites. At the same time, we observe a substantial splitting of the occupied O $2p$ and the empty Ni $e_g$ bands at the surface caused by the effect of surface symmetry breaking (see the top panel of Fig.~\ref{Fig_1}). It leads to a sizeable reduction of the band gap from about 2.6 eV in the quasi-bulk to 2.1 eV at the surface. Moreover, surface relaxation (strain) is seen to result in a small increase of the band gap by less than $\sim$5\%, from 2.0 to 2.1 eV. This result underlines the importance of the crystal-field splitting caused by the effect of surface truncation and relaxations for establishing the Mott insulating state of $(001)$ NiO.

\begin{figure}[t]
\includegraphics[width=0.45\textwidth]{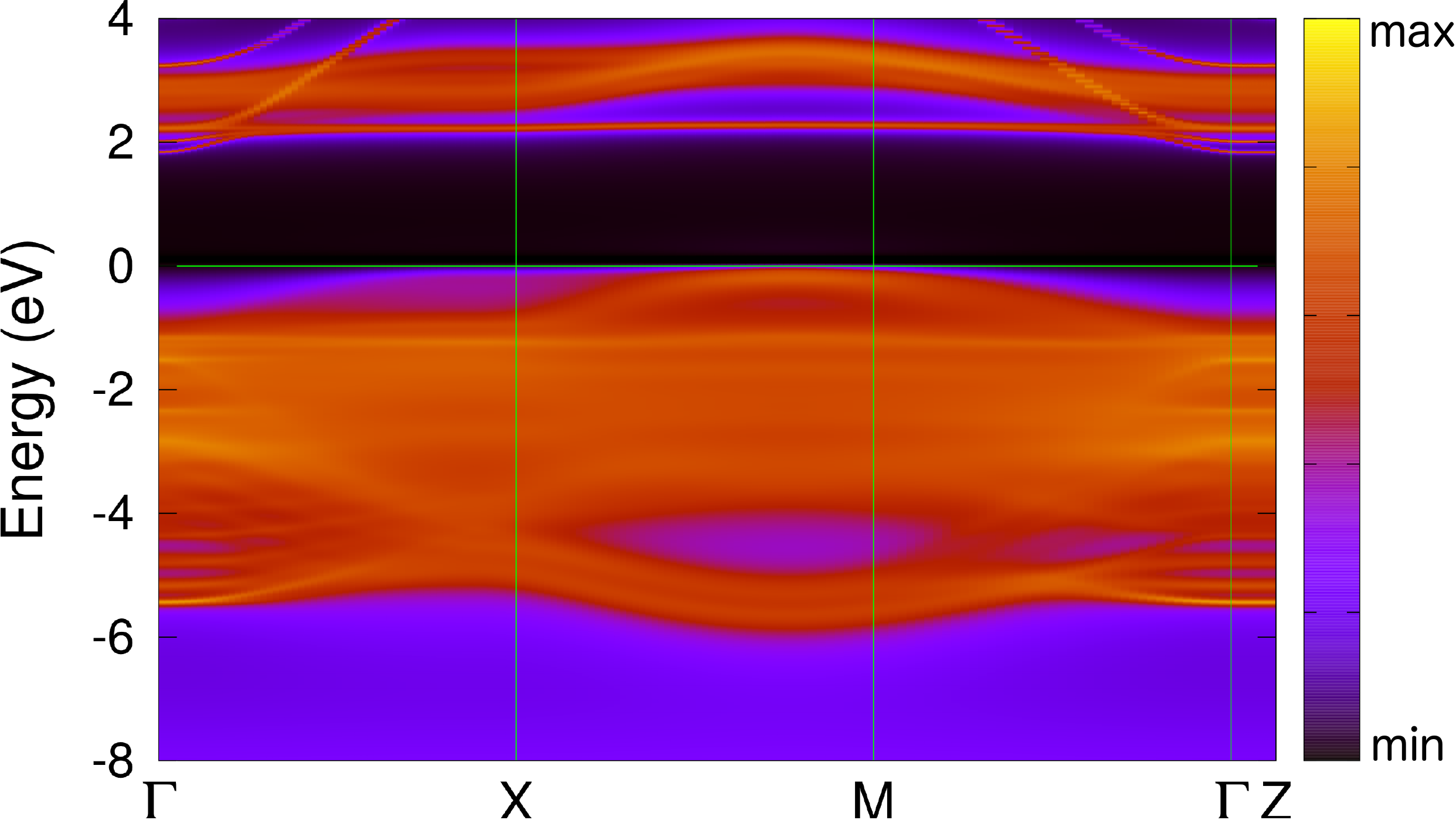}
\caption{(Color online) The {\bf k}-resolved total spectral function 
$A({\bf k},\omega)$ of (001) NiO along the $\Gamma$-X-M-$\Gamma$-Z lines in the Brillouin zone as obtained by DFT+DMFT at a temperature ${T = 390}$~K.}
\label{Fig_2}
\end{figure}

In the surface layer the top of the $(001)$ NiO valence band shows a mixed Ni $x^2-y^2$ and O $2p_z$ character, with a large contribution from the O $2p$ states, caused by the surface symmetry breaking and strain effects. Moreover, in the subsurface ML the top of the valence band has a mixed Ni $3z^2-r^2$ and O $2p_x/p_y$ character. Our results therefore suggest a charge-transfer character of the $(001)$ surface. Moreover, we also notice the importance of orbital-differentiation of the Ni $e_g$ states (for the surface and subsurface NiO-MLs) to characterize the correlated insulating state of $(001)$ NiO. In addition to this, the unoccupied Ni $3z^2-r^2$ surface states are seen to split from the lower edge of the conduction band and form strongly localized (and dispersionless) band at about 2 eV above the $E_F$ (see our results for the {\bf k}-resolved spectra of $(001)$ NiO in Fig.~\ref{Fig_2}). In contrast to this, at the top of the valence band, all bands merge with the continuum of bulk states. 

\begin{figure}[t]
\includegraphics[width=0.45\textwidth]{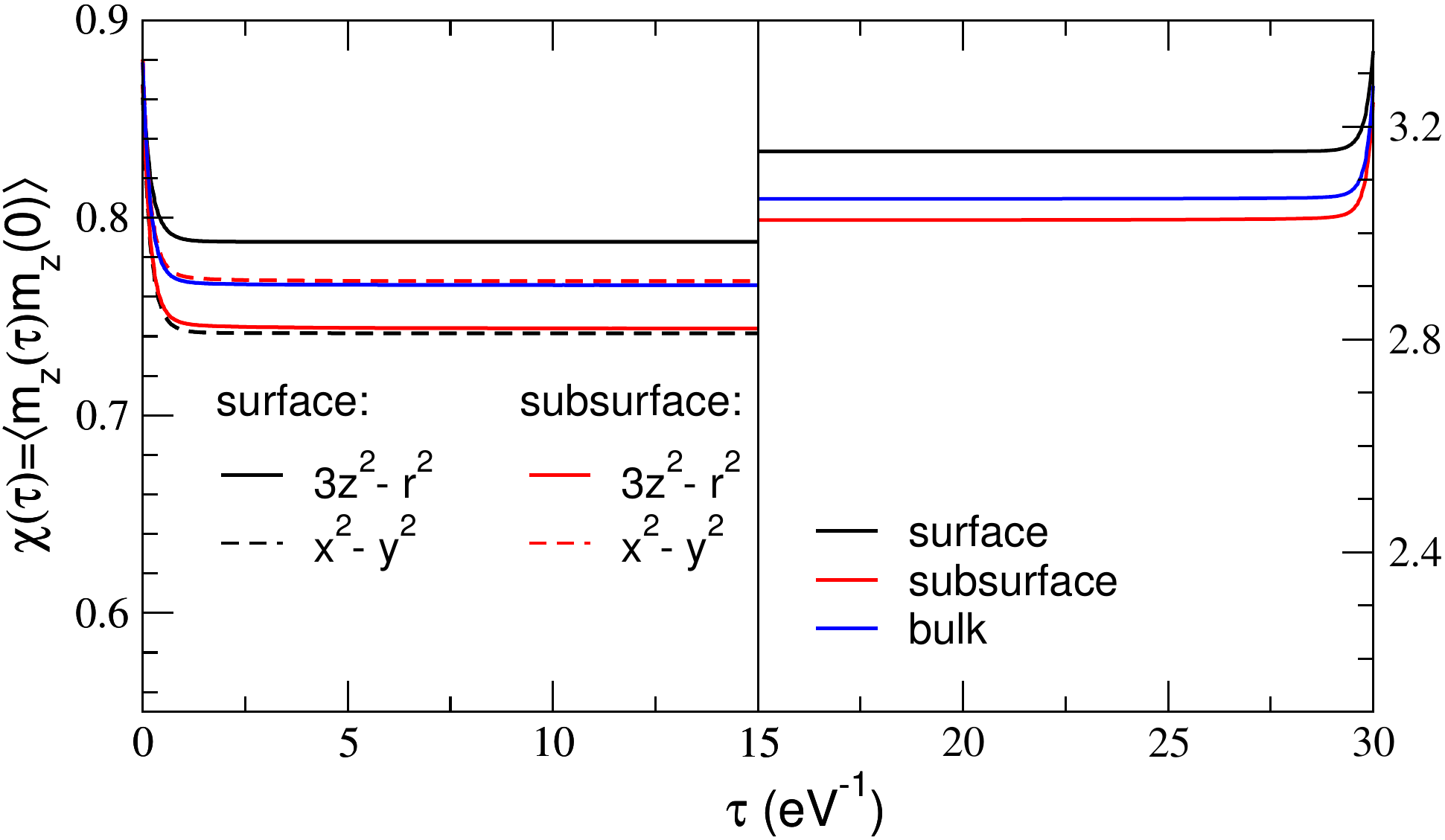}
\caption{(Color online) Local spin-spin correlation function $\chi(\tau)=\langle \hat{m}_z(\tau)\hat{m}_z(0)\rangle$ of paramagnetic $(001)$ NiO as obtained by DFT+DMFT at an inverse temperature $\beta=\frac{1}{k_B T} = 30 eV^{-1}$. Since $\chi(\tau)$ is symmetric with respect to $\beta/2 \pm \tau$, it is plotted, without loss of information, in half the interval from $0$ to $\beta$.
}
\label{Fig_3}
\end{figure}

We also calculated the local (dynamical) susceptibility $\chi(\tau)=\langle \hat{m}_z(\tau)\hat{m}_z(0)\rangle$ for the surface, subsurface, and quasi-bulk Ni$^{2+}$ ions of PM $(001)$ NiO. In Fig.~\ref{Fig_3} we display our results for the N $e_g$ contributions in $\chi(\tau)$. $\tau$ denotes imaginary times.
The value of this quantity around $\tau = \frac{\beta}{2}$ indicates the long time limit, and deviations from this value for small and large $\beta$ indicate strong magnetic fluctuations. In the present case, all the $e_g$ contributions are seen to be almost independent of $\tau$, suggesting that the $3d$ electrons are localized to form fluctuating local moments. In fact, $\chi(\tau)$ is seen to be nearly constant and close to its maximal value $S = 1$ for the Ni $e_g$ states. However, in the surface layer $\chi(\tau)$ is seen to be remarkably larger at all $\tau$, implying higher localization of the surface $e_g$ states. This leads to a higher charge-transfer character of the $(001)$ NiO surface, while for the bulk our results suggest a mixture of a Mott-Hubbard $d$-$d$ and charge transfer character of the band gap  \cite{Sawatzky_1984,Zaanen_1985,Schuler_2005}.

For the $(001)$ NiO we compute the surface energy within DFT+DMFT as $\gamma \simeq (E_{slab}^N - N E_{bulk})/(2S)$ \cite{Singh-Miller_2009}. Here, $E_{slab}^N$ is the total energy of an $N$-formula-unit slab of NiO. $E_{bulk}$ is the total energy of bulk NiO per formula unit. $S$ is the surface area and the factor two accounts for the two surfaces in the slab unit-cell.  In order to minimize numerical differences between the bulk and the slab DFT+DMFT calculations we adopt the same setup parameters within the DFT+DMFT total energy calculations (such as kinetic energy cutoffs, {\bf k}-point sampling, etc.). In particular, we compute the 4-formula-unit-supercell to estimate $E_{bulk}$ for the $(001)$ NiO. Our result 59 meV/\AA$^2$ is in good agreement with previous estimates in the antiferromagnetically ordered phase \cite{Schron_2013,Rohrbach_2004,Bengone_2002}. We note that the nonmagnetic GGA yields 44 meV/\AA$^2$.

\subsection*{b. $(110)$ NiO surface}

We now turn to our results for the PM $(110)$ NiO. To model $(110)$ NiO we design the $(110)$ 11-ML-thick slab with symmetry-equivalent surfaces. It is of 14.59 \AA\ thick, with a vacuum spacer of about 21 \AA. In our DFT+DMFT calculations, we treat the top three NiO MLs as surface, subsurface, and sub-subsurface layers, respectively. The remaining 5-ML-thick layer in the middle is taken as quasi-bulk NiO. We use the four-impurity-site extension of DFT+DMFT to treat the effect of correlations in the $3d$ shell of the structurally distinct Ni ions. In DFT the kinetic energy cutoff for the plane wave basis was set to 50 Ryd for the wave function and 500 Ryd for the charge density. All the calculations are performed in the local basis set determined by diagonalization of the corresponding Ni $3d$ occupation matrices. Structural optimization of the surface and two subsurface $(110)$ NiO layers within FM GGA suggests that surface relaxations in $(110)$ NiO are long-range.
It gives a large inward relaxation of the top layer of about -11.0\% (-0.16 \AA) accompanied by a sizable outward relaxation of the subsurface layer by 3.4\%. The inter-plane relaxation between the first and third layer $\delta_{13}$ is about -3.8\%, while $\delta_{15}\sim -2.9$\%.

\begin{figure}[t]
\includegraphics[width=0.45\textwidth]{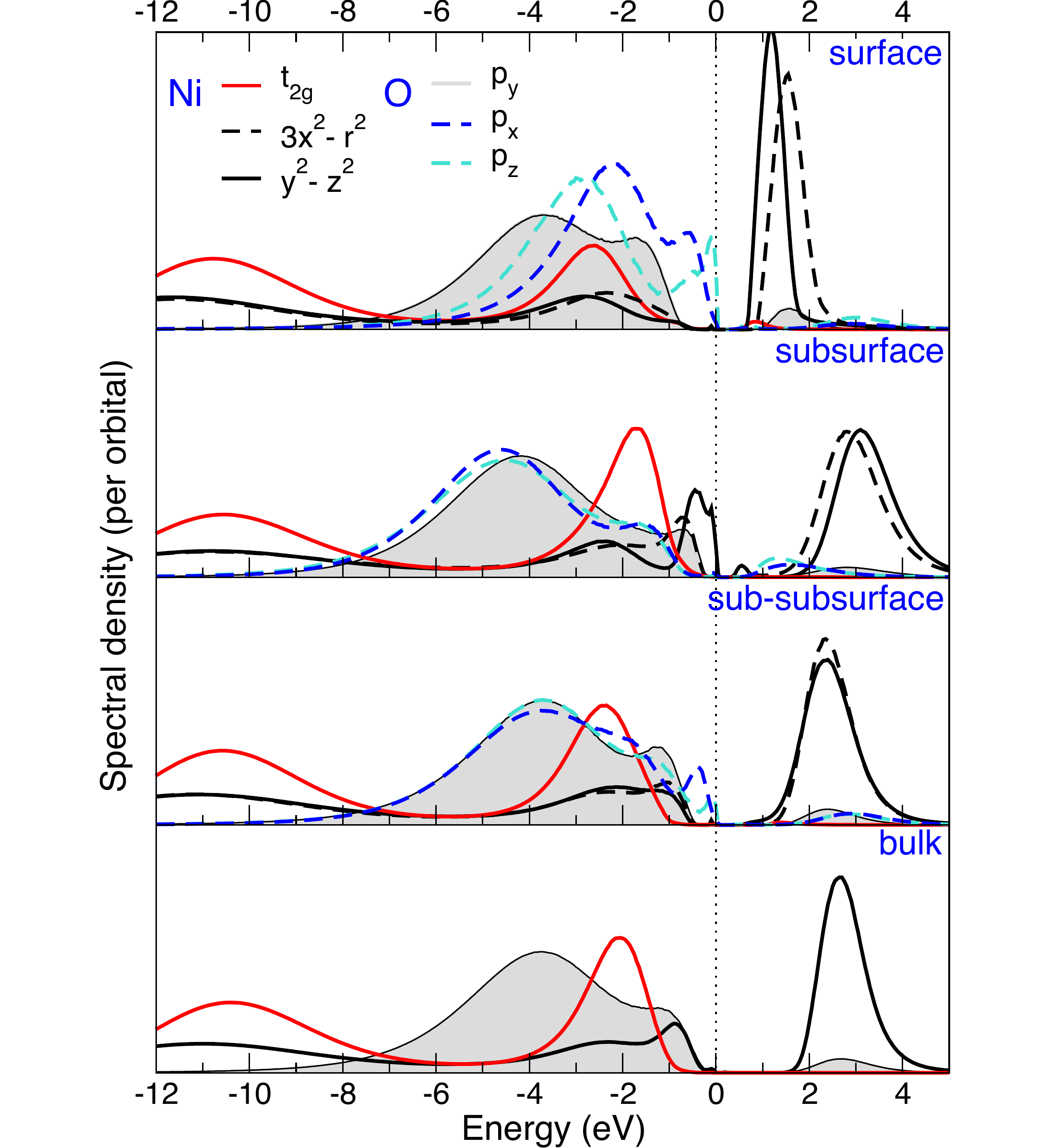}
\caption{(Color online) Ni $e_g$ and O $2p$ spectral functions of paramagnetic $(110)$ NiO obtained by DFT+DMFT at a temperature ${T = 390}$~K.}
\label{Fig_4}
\end{figure}

\begin{figure}[t]
\includegraphics[width=0.45\textwidth]{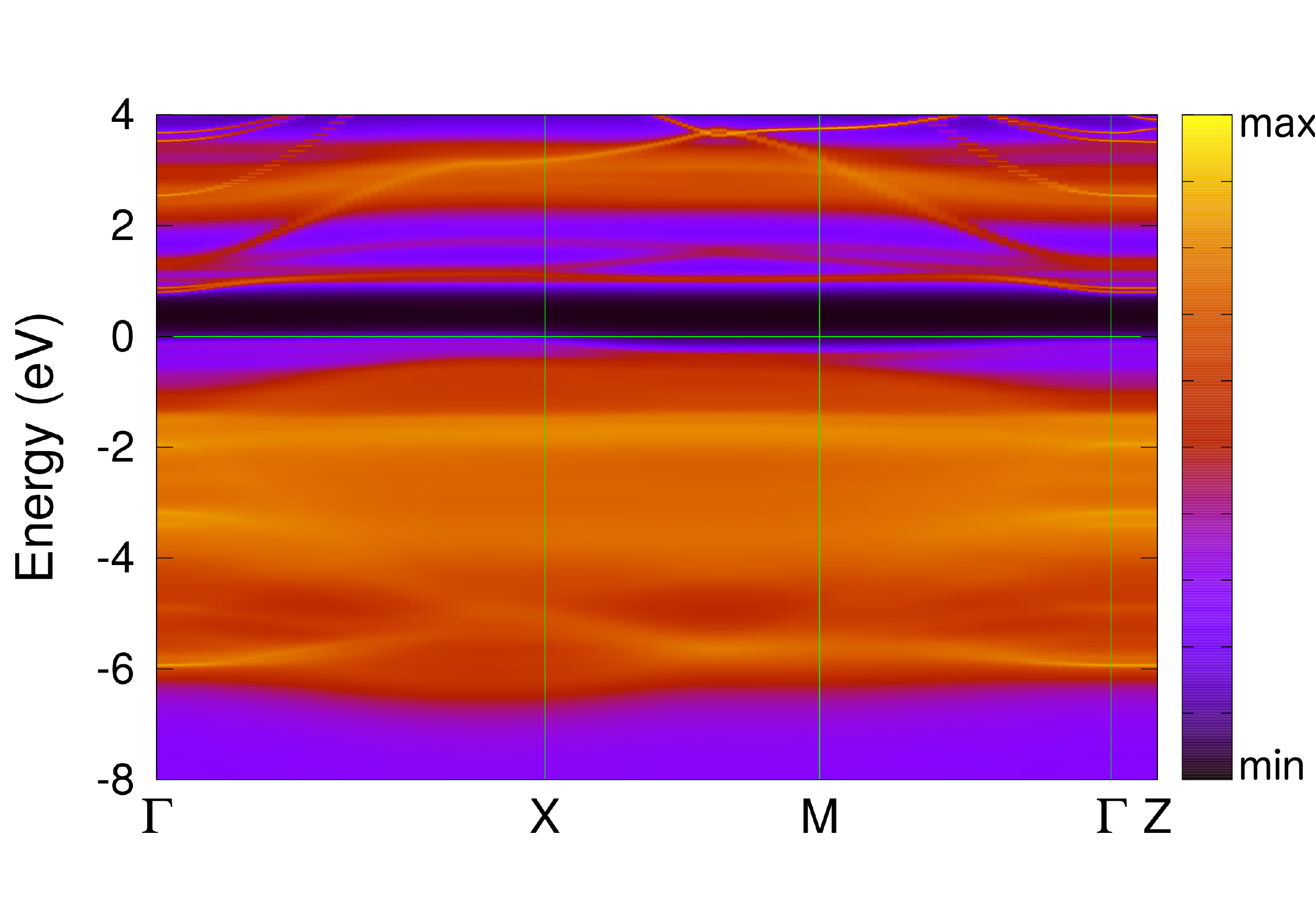}
\caption{(Color online) The {\bf k}-resolved total spectral function 
$A({\bf k},\omega)$ of $(110)$ NiO along the $\Gamma$-X-M-$\Gamma$-Z lines in the Brillouin zone as obtained by DFT+DMFT at a temperature ${T = 390}$~K.}
\label{Fig_5}
\end{figure}

In Fig.~\ref{Fig_4} we summarize our results for the Ni $3d$ and O $2p$ spectral functions of PM $(110)$ NiO. Our results for the {\bf k}-resolved spectra are shown in Fig.~\ref{Fig_5}. We obtain a Mott-Hubbard insulating solution with a relatively large band gap of 0.9 eV. Similarly to $(001)$ NiO, we observe a weak variation of the local moments among the different Ni layers. In particular, the calculated local magnetic moments are 1.83, 1.79, 1.82, and 1.82 $\mu_\mathrm{B}$ for the surface, two subsurfaces, and quasi-bulk Ni-ions, respectively. Our estimate of the fluctuating moments is 1.77, 1.70, 1.73, and 1.75 $\mu_\mathrm{B}$. The Ni $3d$ occupancies are nearly the same for all the Ni sites, differing by less than 0.04. In addition, we observe no sizeable difference between the Ni $3x^2-r^2$ and $y^2-z^2$ orbital occupancies in $(110)$ NiO. The deviations are less than 0.03. Our analysis of the layer-dependent Ni $3d$ and O $2p$ spectral functions reveals a remarkable reduction of the band gap at the surface layer. In fact, it drops by about three times, from 2.9 eV in the quasi-bulk to 0.9 eV in the surface layer, in the relaxed $(110)$ NiO. Most importantly, our result for the (fundamental) gap value of $(110)$ NiO is more than twice smaller than that in the $(001)$ NiO (2.1 eV). In close similarity to $(001)$ NiO, structural relaxations of the surface result in an increase of the band gap from 0.8 to 0.9 eV.

Our results for the Ni $3d$ and O $2p$ spectral functions show a sizeable splitting of the occupied O $2p$ and the empty Ni $e_g$ bands (see Figs.~\ref{Fig_4} and \ref{Fig_5}). While at the top of the valence band, all bands merge with the continuum of bulk states, the unoccupied Ni $e_g$ surface states are seen to split from the lower edge of the conduction band. These Ni $e_g$ states form a localized band with a bandwidth of about 0.4 eV which appears in the bulk band gap at about 1 eV (see Fig.~\ref{Fig_5}). Moreover, a detailed analysis of the top of the valence band reveals a large contribution of the O $2p$ states near the $E_F$, which are strongly mixed with the Ni $3d$ states. It is interesting to note however that the situation significantly differs from that in the $(001)$ NiO or in bulk NiO. In fact, in contrast to the bulk and $(001)$ NiO we observe that the top of the valence bands of $(110)$ NiO has almost pure O $2p$ character at the surface and in the sub-subsurface layer, implying \emph{a pure charge-transfer type} of the band gap of the $(110)$ surface of NiO. In contrast to this, at the subsurface layer, it is nearly purely of Ni $3d$ ($y^2-z^2$) character, suggestive of a Mott-Hubbard type of the band gap in the subsurface layer of $(110)$ NiO. 
Our results therefore document a novel electronic state characterized by an alternating pure charge-transfer and Mott-Hubbard character of the band gap in the surface and subsurface $(110)$ NiO, respectively. This behavior suggests the absence of prominent Zhang-Rice physics at the $(110)$ surface of NiO that reappear in bulk NiO \cite{Kunes_2007,Yin_2008,Leonov_2016}.
We notice that this remarkable alternating charge-transfer and Mott-Hubbard state is robust with respect to the lattice reconstructions of the $(110)$ surface as seen in the spectral function of the bulk-terminated $(110)$ NiO (see Fig.~S2).
This suggests that it is of purely electronic origin.
Moreover, we notice a remarkable orbital-differentiation of the Ni $e_g$ states at the (sub-) surface which seems to be important to characterize the transport properties of $(110)$ NiO. 

\begin{figure}[t]
\includegraphics[width=0.45\textwidth]{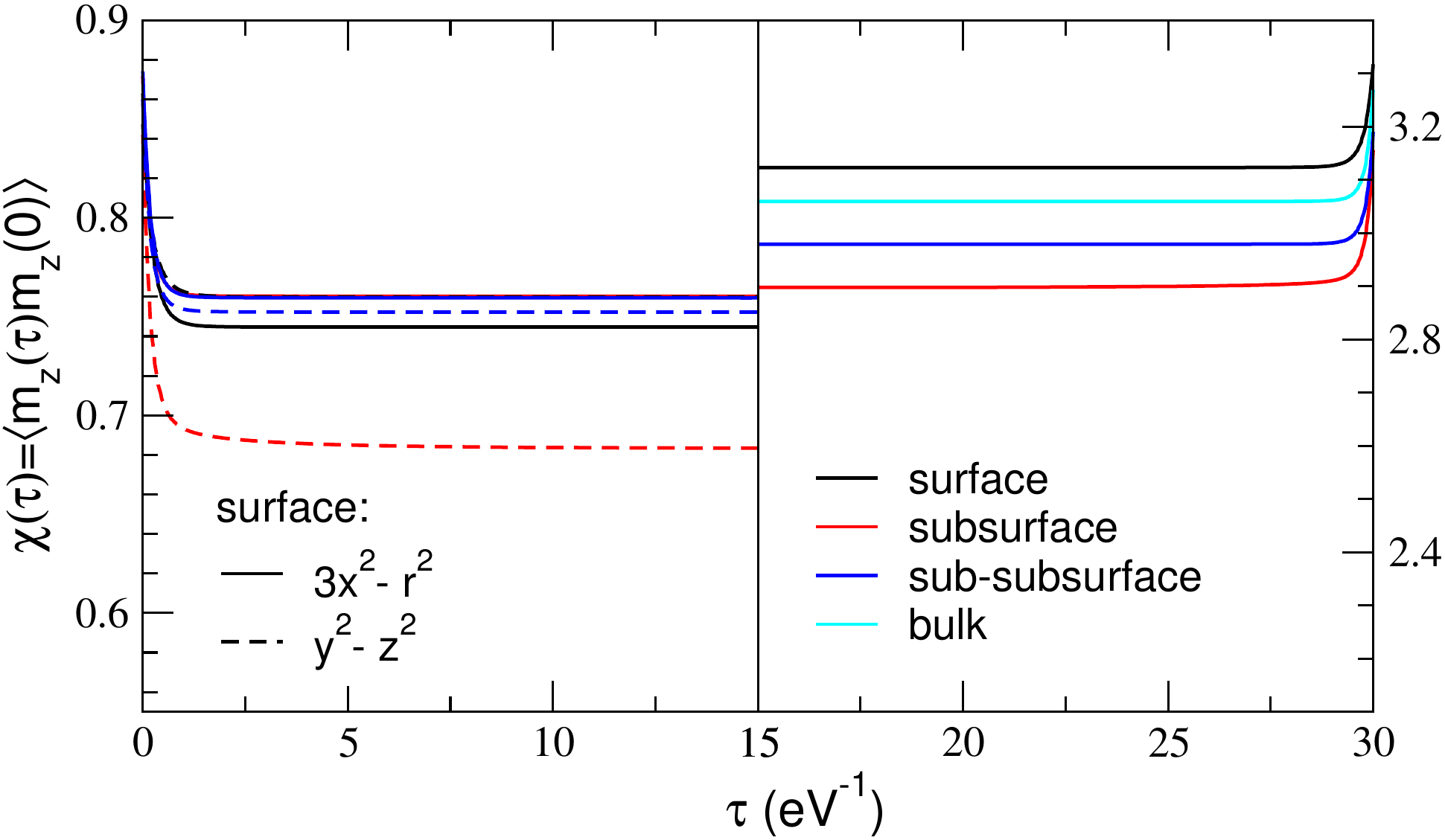}
\caption{(Color online) Local spin-spin correlation function $\chi(\tau)=\langle \hat{m}_z(\tau)\hat{m}_z(0)\rangle$ of paramagnetic $(110)$ NiO as obtained by DFT+DMFT.}
\label{Fig_6}
\end{figure}In Fig.~\ref{Fig_6}{} we display our results for the layer-dependent orbitally-resolved local susceptibility $\chi(\tau)$. We observe that all the $e_g$ contributions are seen to be almost independent of $\tau$, suggesting a strongly localized nature of the Ni $e_g$ states in $(110)$ NiO. Similarly to $(001)$ NiO the surface layer $\chi(\tau)$ is somewhat larger for all $\tau$, implying a higher degree of localization of the surface $e_g$ states. On the other hand, counter-intuitive to this is the reduction of the band gap at the surface layer seen in the Ni $3d$ and O $2p$ spectral function for the $(110)$ NiO surface, which is caused by charge-transfer effects.

We also calculate the surface energy of the relaxed $(110)$ NiO using DFT+DMFT.  In our calculations we use the 8-formula-unit supercell NiO as a reference to estimate $E_{bulk}$. Our result for the surface energy of $(110)$ NiO 125 meV/\AA$^2$ is substantially larger than that for the $(001)$ NiO (59 meV/\AA$^2$), implying that $(001)$ NiO has significantly lower energy facet. The nonmagnetic GGA calculation gives 85 meV/\AA$^2$ for $(110)$ NiO.

\section*{IV. CONCLUSIONS}

In conclusion, we have calculated the electronic structure, magnetic properties, and surfaces energies of the $(001)$ and $(110)$ surfaces of the prototypical correlated insulating material, PM NiO, using a multi-site extension of the DFT+DMFT method. Our results reveal a complex interplay between electronic correlations and surface effects in NiO. We obtain that the electronic structure of the $(001)$ and $(110)$ NiO differs significantly from that of bulk NiO. In both $(001)$ and $(110)$ NiO we observe a sizable reduction of the band gap at the surface. The latter is most significant for the $(110)$ NiO, from 2.9 in the quasi-bulk to 0.9 eV at the surface. Surface relaxations (surface strain) are seen to result in a remarkable increase of the band gap in comparison with the bulk-terminated NiO. 

We observe a substantial splitting of the O $2p$ and Ni $e_g$ bands at the surface caused by the effect of surface symmetry breaking. Our results suggest a charge-transfer character of the $(001)$ and $(110)$ surfaces of NiO. 
Most importantly, for the $(110)$ NiO surface we observe a remarkable electronic state characterized by an alternating pure charge-transfer and Mott-Hubbard character of the band gap in the surface and subsurface NiO layers, respectively. 
This novel form of electronic order stabilized by strong correlations is not driven by lattice reconstructions but of purely electronic origin.
Moreover, we notice the importance of orbital-differentiation of the Ni $e_g$ states to characterize the Mott insulating state of the $(001)$ and $(110)$ NiO. The unoccupied Ni $e_g$ surface states are seen to split from the lower edge of the conduction band, to form strongly localized states in the fundamental gap of bulk NiO. 

Our DFT+DMFT calculations reveal a remarkable difference between the electronic structure of the $(001)$ and $(110)$ NiO surfaces. For example, we obtain a large difference in their (fundamental) band gap value (by about a factor of two) due to the surface effects, $\sim$2.1 eV in the $(001)$ and 0.9 eV for the $(110)$ NiO surfaces. This suggests a higher catalytic activity of the $(110)$ NiO surface than that of the $(001)$ NiO surface. In agreement with previous estimates in the antiferromagnetically ordered phase, our DFT+DMFT calculations yield significantly different surface energies for the $(001)$ and $(110)$ NiO surfaces, of 59 and 125 meV/\AA$^2$, respectively. Thus, the $(001)$ NiO surface is found to have a significantly lower energy facet. 
We find that the effect of electron-electron correlations in NiO results in a sizable enhancement of the NiO surface energies by 34-48\% with respect to the nonmagnetic GGA.
Overall, our results for the electronic structure, magnetic properties, and surface energies of $(001)$ and $(110)$ NiO agree well with experimental data measured in the antiferromagnetic phase.
Our results further reveal an intriguing independence of properties such as the energetical hierarchy of different surfaces of the presence or absence of antiferromagnetic order.

In our study we demonstrate the ability of multi-site DFT+DMFT to compute the properties of transition metal oxides surfaces. The DFT+DMFT computations can be used to promote our understanding of the physics of surfaces, which are promising for applications. Further studies could for example deal with NiO deposited on SrTiO$_3$, elucidating the increased reactivity with water of such systems \cite{diebold} or the design of highly active catalysts, e.g., for developing hybrid nickel-zinc and zinc-air batteries \cite{Lee_2016}. NiO is also being actively investigated for its resistive switching properties \cite{resistive-switching,RS2,RS3} for the description of which extensions of DFT+DMFT might also  prove useful.

\begin{acknowledgments}

We thank S.~Backes, S.~Panda, D. D.~Sarma, and I. A. Abrikosov for valuable discussions. Theoretical analysis of the electronic properties of (001) and (110) NiO was supported by the state assignment of Minobrnauki of Russia (theme ``Electron'' No. AAAA-A18-118020190098-5). The DFT+DMFT calculations of magnetic properties of (001) and (110) NiO were supported by Russian Science Foundation (project No. 19-72-30043). S.B. acknowledges support from the European Research Council under grant agreement 617196, project CORRELMAT, and from IDRIS/GENCI Orsay under project t2020091393.

\end{acknowledgments}

\pagebreak
\hfill
\pagebreak

\section*{Supplementary Material}
\beginsupplement

In this supplementary material, we display additional information
concerning the spectral properties of the investigated NiO surfaces.

\begin{figure}[h]
\includegraphics[width=0.5\textwidth]{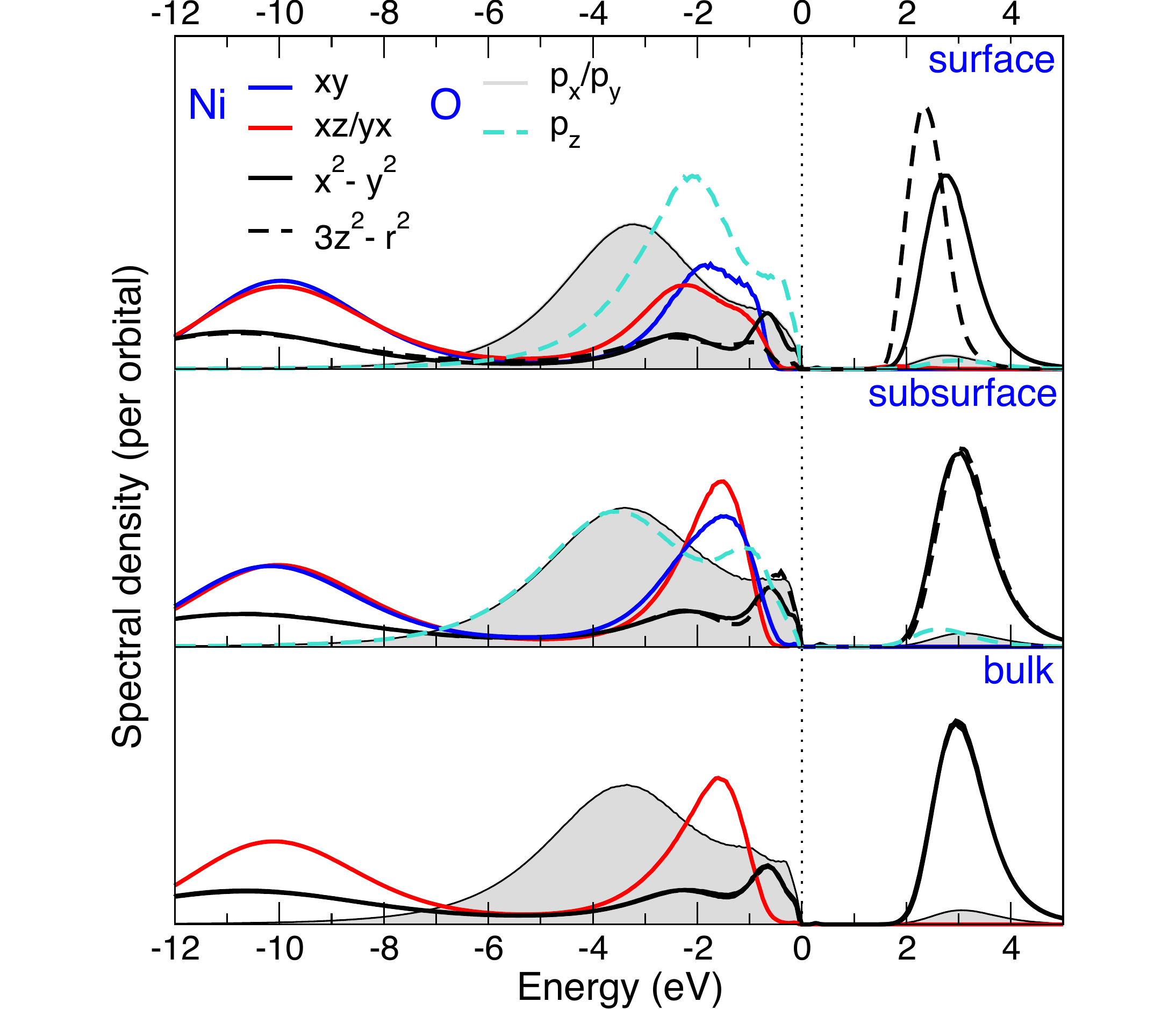}
\caption{(Color online) Ni $3d$ and O $2p$ spectral functions calculated by DFT+DMFT for the bulk-terminated $(001)$ NiO at a temperature $T=390$~K.}
\label{Fig_S1}
\end{figure}

\begin{figure}[h]
\includegraphics[width=0.5\textwidth]{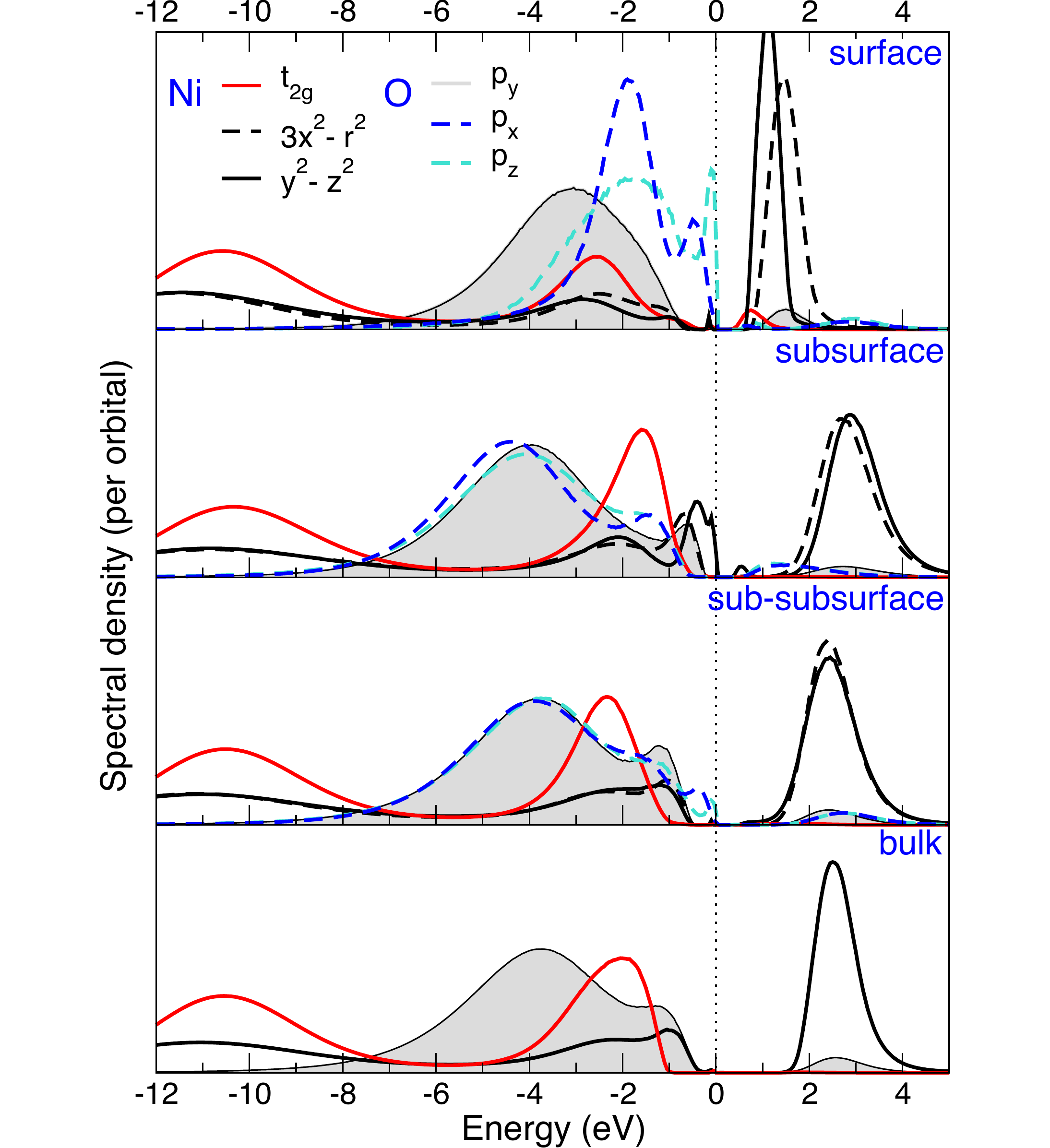}
\caption{(Color online) Ni $3d$ and O $2p$ spectral functions calculated by DFT+DMFT for the bulk-terminated $(110)$ NiO at a temperature $T=390$~K.}
\label{Fig_S2}
\end{figure}

\end{document}